\documentclass[acmsmall, screen, review=false, anonymous=false]{acmart}

\usepackage{amsmath,amsfonts}
\usepackage{algorithmic}
\usepackage{graphicx}
\usepackage{textcomp}
\usepackage{xcolor}
\usepackage{lscape}
\usepackage{pifont}
\usepackage{tcolorbox}
\usepackage{graphics}
\usepackage{booktabs}
\usepackage{algorithmic}
\usepackage{graphicx}
\usepackage{textcomp}
\usepackage{xcolor}
\usepackage{wasysym}
\usepackage{multirow}
\usepackage{makecell}
\usepackage{subcaption}
\usepackage{enumitem}
\usepackage{listings}
\usepackage{soul,color}
\usepackage{subcaption}
\usepackage{soul}
\usepackage{tikz}
\usepackage[framemethod=TikZ]{mdframed}
\usepackage{wrapfig}
\usepackage{flushend}
\usepackage{tablefootnote}
\usepackage{blindtext}
\usepackage{float}
\usepackage{lipsum}
\usepackage{titlesec}
\usepackage{url}
\usepackage[utf8]{inputenc}
\usepackage{multirow}
\usepackage{colortbl}

\usetikzlibrary{shapes,arrows}
\usetikzlibrary{positioning, shapes.geometric, arrows.meta}

\newcommand{\paratitle}[1]{
    \setlength{\parskip}{0.3\baselineskip}  \noindent\textbf{#1}%
}

\newcommand{\noMargine}{\setlength{\parskip}{0pt}}

\newcommand{\quoteFrame}[1]{%
    \vspace{0.3em} %
    \begin{mdframed}[leftline=true, topline=false, bottomline=false, rightline=false, linewidth=2pt, linecolor=gray, innertopmargin=2pt, innerbottommargin=2pt] 
       #1
    \end{mdframed}
    \vspace{-4pt}
}

\newcommand{\find}[1]{
    \begin{tcolorbox}[leftrule=1mm,toprule=0mm,bottomrule=0mm,left=1pt,right=2pt,top=2pt,bottom=2pt
    ]
    \em #1
    \end{tcolorbox}
}

\newcommand{\researchQuestion}[3]{
  \noindent\textbf{#1} {#2}
  \hangindent=2.5em \hangafter=1
  \setlength{\parskip}{0pt} 
  \label{#3} 
}

\AtBeginDocument{%
  }

\setcopyright{none}
\renewcommand\footnotetextcopyrightpermission[1]{} 

\begin{document}

\title{\textit{``I Don’t Use AI for Everything''}: \\ Exploring Utility, Attitude, and Responsibility of AI-empowered Tools in Software Development}

\makeatletter
\let\@authorsaddresses\@empty
\makeatother
\author{Shidong Pan}
\affiliation{
  \institution{CSIRO's Data61 \& Australian National University}
  \country{Australia}
  }
\author{Litian Wang}
\affiliation{
  \institution{Australian National University}
  \country{Australia}
  }
\author{Tianyi Zhang}
\affiliation{
  \institution{Australian National University}
  \country{Australia}
  }
\author{Zhenchang Xing}
\affiliation{
  \institution{CSIRO's Data61 \& Australian National University}
  \country{Australia}
  }
\author{Yanjie Zhao}
\affiliation{
  \institution{Huazhong University of Science and Technology}
  \country{China}
  }
\author{Qinghua Lu}
\affiliation{
  \institution{CSIRO's Data61}
  \country{Australia}
  }
\author{Xiaoyu Sun}
\authornote{Corresponding author, Xiaoyu.Sun1@anu.edu.au}
\affiliation{
  \institution{Australian National University}
  \country{Australia}
  }

\renewcommand{\shortauthors}{Pan et al.}

\begin{abstract}
    \textbf{[Abstract]} AI-empowered tools have emerged as a transformative force, fundamentally reshaping the software development industry and promising far-reaching impacts across diverse sectors.
    This study investigates the adoption, impact, and security considerations of AI-empowered tools in the software development process. Through semi-structured interviews with 19 software practitioners from diverse backgrounds, we explore three key aspects: the utility of AI tools, developers' attitudes towards them, and security and privacy responsibilities. Our findings reveal widespread adoption of AI tools across various stages of software development. Developers generally express positive attitudes towards AI, viewing it as an efficiency-enhancing assistant rather than a job replacement threat. 
    However, they also recognized limitations in AI's ability to handle complex, unfamiliar, or highly specialized tasks in software development.
    Regarding security and privacy, we found varying levels of risk awareness among developers, with larger companies implementing more comprehensive risk management strategies.
    Our study provides insights into the current state of AI adoption in software development and offers recommendations for practitioners, organizations, AI providers, and regulatory bodies to effectively navigate the integration of AI in the software industry.
\end{abstract}

\maketitle

\section{Introduction}
In recent years, generative AI (GAI) has emerged as a transformative force, fundamentally reshaping software development and promising far-reaching impacts across diverse sectors. According to OpenAI President Greg Brockman~\cite{chatgpt}, within just five days of ChatGPT's initial public release on November 30, 2022, more than one million users had registered. One recent research estimated that GAI could add between \$2.6 trillion to \$4.4 trillion annually to the global economy~\cite{chui2023economic}. Besides, researchers anticipate GAI's significant impact to extend across a broad spectrum of industries, including but not limited to manufacturing, marketing, banking, healthcare, and education~\cite{chui2023economic, mondal2023bell, ooi2023potential,brynjolfsson2023generative,zhang2023test}. This convergence of rapid adoption, enormous economic potential, and cross-sector applicability underscore the critical importance of understanding GAI's integration and implications, particularly in the realm of software development where its influence is already profoundly felt~\cite{ebert2023generative}.

Although these AI-empowered tools\footnote{In this paper, we use the term AI tools to refer to AI-empowered tools.} have been widely recognized for their potential in various areas we mentioned above, the utility of these AI tools in the software development process has not been completely investigated. As these tools become increasingly integrated into software development, there is a necessary need for rigorous academic research to provide a comprehensive understanding of AI usage in the software development process. Also,  Banh \& Strobel~\cite{banh2023generative} pointed out that despite AI's potential for transformative capabilities, it still faces challenges similar to those encountered by traditional machine learning systems. These potential and challenges highlight the importance of understanding how developers interact with AI tools. Therefore, we aim to further investigate developers' attitudes towards the AI tools they use, exploring whether their perceptions and actual experiences differ. At the same time, Si et al.~\cite{si2024solution} raised concerns that in the process of using GAI, developers tend to ignore private information. McGraw \& Inc~\cite{1281254} also illustrated the importance of software security in building software. This means that while developers are increasingly using AI tools to assist in their daily work, the associated issues of reliability, as well as possible security and privacy (S\&P) risks, should also be considered. So we wonder how developers consider security and privacy responsibility in the development process.



Therefore, as AI tools play an increasingly important role in the software development process~\cite{solohubov2023accelerating,charankar2024microservices,ozkaya2023next}, it is crucial to understand the three key aspects of utility, attitude, and responsibility as aforementioned. 
We found that while current research has explored AI tool usage~\cite{ozpolat2023artificial}, its impact~\cite{rajbhoj2024accelerating}, challenges~\cite{harman2012role}, and potential risks~\cite{klemmer2024using} from various perspectives, most studies tend to focus on a specific aspect, lacking a comprehensive summary and overview of all relevant factors. Additionally, existing research has primarily focused on analyzing the impact of AI on software development during the integration process, by a series of literature reviews\cite{nguyen2023generative, martinez2022software, fi15060192}. 
Few studies have focused on developers' views, and research on specific guidelines and ethical challenges is lacking. This indicates a need for future research to take a more holistic approach to examining the multidimensional effects of AI tools in real-world applications to better inform both developers and industry practices.

\renewcommand{\thefootnote}{\arabic{footnote}}

Therefore, our research aims to gain insights into developers' experiences of using AI tools and their considerations of responsible use through semi-structured interviews, focusing on how software developers interact with AI tools in their daily work. 
The interview process will refer to the process of previous research to ensure that our research effectively builds on existing knowledge.\cite{fourne2023s, wermke2023always}. We seek answers to the following research questions:

\phantomsection
\researchQuestion {RQ1~}{``\textit{What roles do AI tools play in software development?}'' We want to comprehensively understand AI tools and how they work in the software development process. Therefore, we will focus on which kind of AI tools are mostly used and how developers integrate AI tools into various stages of software development. This question will also explore how these tools have been adopted in the software development process.}{rq:one}

\phantomsection
\researchQuestion{RQ2~}{``\textit{What are the attitudes and perspectives of developers regarding the application of AI tools to the software development process?}'' We aim to gain an in-depth understanding of the attitudes and perceptions of developers towards the AI tools they use, not only their acceptance of these tools but also their perceptions of the positive impacts and potential challenges that AI can bring. We are also interested in their concerns regarding job displacement.}{rq:two}

\phantomsection
\researchQuestion{RQ3~}{``\textit{What potential S\&P risks are developers considering when using AI-empowered tools?}'' We are interested in the potential risks of using AI in software development to understand the risks developers weigh when integrating these tools into their workflows. We will also explore whether developers and companies are taking steps to anticipate and mitigate these risks, and their perceptions toward mitigation measurements and regulations.}{rq:three}

By answering these questions, we hope to provide a comprehensive overview of AI tools' status quo in software development, evaluating their impact on productivity and the importance of responsible use. We will also develop specific recommendations and guidelines to assist stakeholders in effectively utilizing these tools.

As for the utility of AI tools, we found that almost all participants have integrated AI tools into their daily workflows. AI tools are now involved in almost every stage of software development, being applied to various tasks and taking on diverse roles (Section~\ref{sec_utility}).
As for the attitude, most developers hold a positive outlook on AI. AI tools have not fundamentally changed developers' basic workflows but rather introduced numerous positive impacts. It also comes with many limitations so there are no job replacement concerns (Section~\ref{sec_attitude}). Regarding responsibility, most developers possess a strong sense of risk management. While some developers may not be particularly concerned about these risks, this often stems from their companies' comprehensive management measures. The level of risk management typically correlates with company size and type. Larger commercial organizations implementing more stringent practices (Section~\ref{sec_responsibility}).
We further discuss the broader adoption of AI tools in the industry, exploring strategies such as prompt engineering and outcome evaluation, as well as future perspectives. Additionally, we provide targeted recommendations for stakeholders, and regulatory bodies on effectively addressing the widespread use of AI (Section~\ref{sec_discuss}). We believe this work could significantly contribute to the community's understanding of AI tool adoption in the software development ecosystem, offering valuable insights and recommendations for both practitioners and researchers.


\section{Methodology}\label{sec_method}

To fully understand the growing impact of AI tools in software development and the status quo across the industry, we conducted an interview-based study in this paper. This qualitative approach allows for an in-depth exploration of practitioners' experiences and perspectives~\cite{turner2010qualitative, merriam2015qualitative}, crucial for comprehending the nuanced realities of AI adoption throughout the software development life cycle. We show the specific processes of our research in Figure 1.
Notably, this study received approval from the Institutional Review Board (IRB). For more details, please refer to the artifact repository.




\setlength{\belowcaptionskip}{1pt} 
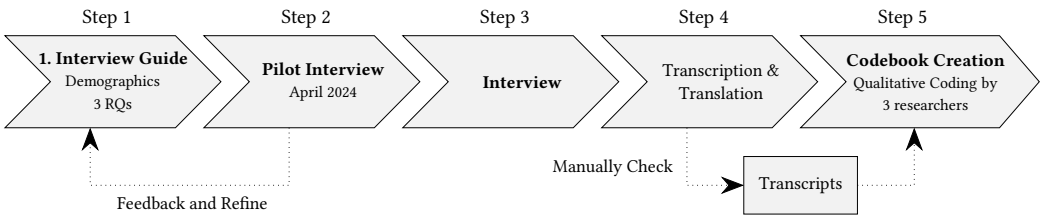
\begin{figure}[h]
    \centering
    \resizebox{\textwidth}{!}{
    \begin{tikzpicture}[node distance=1.5cm]
    \draw[fill=gray!10] 
        (-0.2, 0) -- (-1, 0.8) -- (2, 0.8) -- (2.8, 0) -- (2, -0.8) -- (-1, -0.8) -- cycle; 
    \node[align=center, font=\small] at (0.9, 0) {\textbf{1. Interview Guide}\\ \footnotesize{Demographics} \\ \footnotesize{3 RQs} };
    \node[align = center] at (0.8, 1.1) {Step 1};
    
    \begin{scope}[xshift=3.5cm] 
    \draw[fill=gray!10] 
        (-0.2, 0) -- (-1, 0.8) -- (2, 0.8) -- (2.8, 0) -- (2, -0.8) -- (-1, -0.8) -- cycle;
    \node[align=center, font=\small] at (1.1, 0) {\textbf{Pilot Interview} \\ \footnotesize{April 2024}};
    \node[align = center] at (0.8, 1.1) {Step 2};
    \end{scope}

    \begin{scope}[xshift=7cm] 
    \draw[fill=gray!10] 
        (-0.2, 0) -- (-1, 0.8) -- (2, 0.8) -- (2.8, 0) -- (2, -0.8) -- (-1, -0.8) -- cycle;
    \node[align=center, font=\small] at (1.1, 0) {\textbf{Interview}};
     \node[align = center] at (0.8, 1.1) {Step 3};
    \end{scope}

    \begin{scope}[xshift=10.5cm] 
    \draw[fill=gray!10] 
        (-0.2, 0) -- (-1, 0.8) -- (2, 0.8) -- (2.8, 0) -- (2, -0.8) -- (-1, -0.8) -- cycle;
    \node[align=center, font=\small] at (1.1, 0) {Transcription \& \\Translation};
     \node[align = center] at (0.8, 1.1) {Step 4};
    \end{scope}

    \begin{scope}[xshift=12cm] 
    \draw[fill=gray!10] 
        (0, -1.3) -- (2, -1.3) -- (2, -2.3) -- (0, -2.3) -- cycle;
    \node[align=center, font=\small] at (1, -1.8) {Transcripts};

    \end{scope}

    \begin{scope}[xshift=14cm] 
    \draw[fill=gray!10] 
        (-0.2, 0) -- (-1, 0.8) -- (2.5, 0.8) -- (3.2, 0) -- (2.5, -0.8) -- (-1, -0.8) -- cycle;
    \node[align=center, font=\small] at (1.2, 0) {\textbf{Codebook Creation} \\ \footnotesize{Qualitative Coding by} \\ \footnotesize{3 researchers}};
     \node[align = center] at (0.8, 1.1) {Step 5};
    \end{scope}

    \draw[dotted, -{Stealth[width=3mm,length=4mm]}] 
        (4, -0.8) -- ++(0,-1) -- ++(-3.5, 0) -- ++(0, 1) ;
    \node[align=center, font=\small] at (2.3, -2.1) {Feedback and Refine};

     \draw[dotted, -{Stealth[width=3mm,length=4mm]}] 
        (11, -0.8) -- ++(0,-1) -- ++(1, 0) ;
    \node[align=center, font=\small] at (9.7, -1.5) {Manually Check};

    \draw[dotted, -{Stealth[width=3mm,length=4mm]}] 
        (14, -1.8) -- ++(1,0) -- ++(0, 1) ;
    
    \end{tikzpicture}
    }
    \caption{Research Process}
    \label{fig:enter-label}
\end{figure}






  


 
\subsection{Participant Recruitment}


We recruited participants for our study through a combination of professional networks and targeted recruiting, which is a common recruitment method in interview-based study~\cite{newman2009recruitment, wermke2023always}. 
Our recruitment strategy focused on professionals from our own network, specifically targeting individuals working in the software industry across various roles, geographical locations, company sizes, and levels of work experience. 
In total, we conducted interviews with 19 participants between April and August 2024. We ensured all participants were professionals in software-related fields and aimed for diversity in terms of geographic representation—including participants from China, the United States, Australia, and Europe—as well as in professional roles. Due to our targeted recruitment approach, the sole eligibility criterion was participants' experience in utilizing AI tools in software development. As compensation for their valuable time as domain experts, we offered each participant \$30 or the equivalent value in local Amazon giftcards. Table~\ref{tab_demographics} provides an overview of the interviewed participants' demographics.

\begin{table}[t]
\caption{Demographics of 19 participants. $^1$Position abbreviation: SDE means software development engineer, MLE means machine learning engineer, PM means project manager.}

\label{tab_demographics}
\resizebox{\textwidth}{!}{
\begin{tabular}{@{}ccccllcl@{}}
\toprule
 & \multicolumn{7}{c}{Participants Technical Profile} \\ \cmidrule(l){2-8} 
\multirow{-2}{*}{Alias} & \multicolumn{1}{c}{Country} & \multicolumn{1}{c}{Exp. of SDE } & \multicolumn{1}{c}{Exp. of AI Tools} & Position & \multicolumn{1}{l}{Area} & \multicolumn{1}{c}{Gender} & \multicolumn{1}{l}{Software Stack}\\ \midrule
P01 & Singapore & 6-7 months & 1-2 years & SDE$^1$ & Software Development & Male & C \#, SQL server\\
\rowcolor[HTML]{EFEFEF} 
P02 & China & 9-10 years & 1-2 years & Co-founder & Web2, Web3 & Male & Java, Python, Go, Vue \\
P03 & America & 6 years & 1-2 years & MLE$^1$& Machine Learning & Male & TensorFlow, Python   \\
\rowcolor[HTML]{EFEFEF} 
P04 & China & 6 years & 1-2 years & Game Designer \& PM$^1$& Computier Graphic & Female & C++   \\
P05 & Singapore & 3 months & 1-2 years & DevOps Engineer & DevOps & Male & Java, JS, C\#  \\
\rowcolor[HTML]{EFEFEF} 
P06 & China & 6 years & 1-2 years & Analyst Programmer& Software Development& Male & Java, C \#, Python \\
P07 & Singapore & 1 year & 1-2 years & SDE$^1$& Mobile App Development & Male & Flutter, Java  \\
\rowcolor[HTML]{EFEFEF} 
P08 & China & 2-3 years & 1-2 years & Senior MLE$^1$& Algorithm \& AI & Male & C, Python, Shell \\
P09 & Australia & 7-8 years & 1-2 years & Software Programmer & Research Engineer & Male & JavaScript, Python \\
\rowcolor[HTML]{EFEFEF} 
P10 & America & 3 years & 0.5 years & SDE$^1$& Software Development & Male & C\#, Python, Go  \\
P11 & Hungary & 5 years & 0.5-1year& DevOps Engineer & DevOps and Cloud Computing& Male & Java, Python, Shell \\
\rowcolor[HTML]{EFEFEF} 
P12 & China & 13 years & 1-2 years & Senior Technical Expert & Software Development & Male & Java \\
P13 & China & 8 years &  1 year& PM$^1$& Software Development & Male & Python  \\
\rowcolor[HTML]{EFEFEF} 
P14 & China & 10 years & 1-2 years & Senior SDE$^1$& Software Development & Male & Java \\
P15 & America & 6 years & 1-2 years& Senior SDE$^1$& Autopilot & Female & Python, C++ \\
\rowcolor[HTML]{EFEFEF} 
P16 & Netherland &  5 years& 1-2 years & SDE$^1$&  Software Development& Female &  Scala, Grails, Angular\\
P17 & America & 1 years & 1 year & Algorithm Enigneer & LLM \& SE & Male & Python, Java, C++  \\
\rowcolor[HTML]{EFEFEF} 
P18 & Australia & 1 years & 0.5 years & SDE$^1$&  Software Development& Male &   -\\
P19 & America & 3 years&  2 years&  Senior Researcher &  -& Male &    C++, Python\\ \bottomrule
\end{tabular}
}%
\vspace{-8pt}
\label{tab:my-table}
\end{table}
\subsection{Interview Procedure}
We conducted 19 interviews using a lead interviewer and a backup interviewer configuration. 
This approach is widely adopted in interview-based research, as seen in previous studies~\cite{fourne2023s, wermke2023always, hielscher2023employees}. This setup allowed the lead interviewer to focus on asking questions and actively listening to the interviewee, while the backup interviewer ensured comprehensive coverage of all questions, asked additional follow-up questions as needed, and could take over in case of any connection issues.

Our interview guide was developed based on the research questions and refined through discussions with researchers on our team. 
To further enhance the guide, we conducted a pilot interview with team members, lasting approximately 23 minutes. Based on the feedback received, we made minor adjustments, including the addition of follow-up questions, refinement of wording, and optimization of time allocation to improve the interview flow. Subsequently, we commenced the formal interview process.

All interviews were conducted remotely. We recommended Zoom as the primary platform but accommodated participants' preferences for alternatives like Microsoft Teams or VooV Meeting. We advertised the interviews to last 30-40 minutes but scheduled hour-long appointments to allow for flexibility. The median duration of the actual interview portion, excluding introductions, explanation of consent forms, and debriefing, was 32:25 minutes, with the shortest interview lasting 22:15 minutes and the longest extending to 47:34 minutes. 

Our methodology centers on non-leading, open-ended questions to encourage interviewees to elaborate freely on their experiences and thoughts. Each interview section began with a general question, allowing participants to express their initial perspectives. More specific follow-up questions were only asked if certain points were not addressed organically. All interviewers were instructed to avoid priming participants and to maintain a non-judgmental stance.

\subsection{Interview Structure}
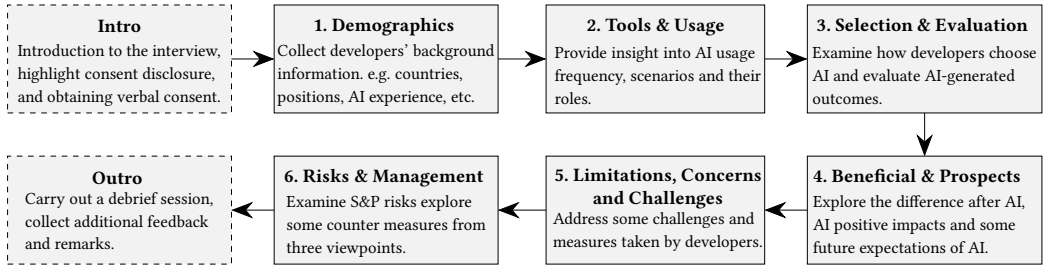
\begin{figure}[ht]
    \centering
    \resizebox{\textwidth}{!}{
    \begin{tikzpicture}[node distance=1.5cm]
    
    \draw[dashed, fill=gray!10] 
        (-1.2, 1) -- (3, 1) -- (3, -1) -- (-1.2, -1) -- cycle; 
    \node[align =center] at (0.9,0.6){\textbf{Intro}};
    \node[align=left] at (0.9, -0.3) {\small{Introduction to the interview,} \\ \small{highlight consent disclosure,} \\ \small{and obtaining verbal consent. }};
    
    \begin{scope}[xshift=5cm] 
    \draw[fill=gray!10] 
           (-1.2, 1) -- (3, 1) -- (3, -1) -- (-1.2, -1) -- cycle; 
           
    \node[align = center] at (0.9,0.6){
    \textbf{1. Demographics} 
    };
    
    \node[align=left] at (0.9, -0.3) { \small{Collect developers' background} \\ \small{information. e.g. countries,} \\ \small{positions, AI experience, etc.}
    };
    \end{scope}

    \begin{scope}[xshift=10cm] 
    \draw[fill=gray!10] 
            (-1.1, 1) -- (3, 1) -- (3, -1) -- (-1.1, -1) -- cycle;
    \node[align = center] at (1,0.6){
    \textbf{2. Tools \& Usage} 
    };
    \node[align=left] at (1, -0.3) { \small{Provide insight into AI usage} \\ \small{frequency, scenarios and their} \\ \small{roles.}} ;
    \end{scope}

    \begin{scope}[xshift=15cm] 
    \draw[fill=gray!10] 
            (-1.2, 1) -- (3.2, 1) -- (3.2, -1) -- (-1.2, -1) -- cycle; 
    \node[align = center] at (1,0.65){
    \textbf{3. Selection \& Evaluation} 
    };
    \node[align=left] at (1, -0.3) { \small{Examine how developers choose} \\ \small{AI and evaluate AI-generated} \\\small{outcomes.}} ;
    \end{scope}

    \begin{scope}[xshift=15cm,yshift = -2.8cm] 
    \draw[fill=gray!10] 
            (-1.2, 1) -- (3.2, 1) -- (3.2, -1) -- (-1.2, -1) -- cycle; 
    \node[align=center] at (0.9,0.6){
    \textbf{4. Beneficial \& Prospects} 
    };
    \node[align=left] at (0.9, -0.3) { \small{Explore the difference after AI,} \\ \small{AI positive impacts and some} \\ \small{future expectations of AI.} };
  
    \end{scope}

    \begin{scope}[xshift=10cm,yshift = -2.8cm] 
    \draw[fill=gray!10] 
            (-1.1, 1) -- (3, 1) -- (3, -1) -- (-1.1, -1) -- cycle; 
    \node[align=center] at (1,0.4){
    \textbf{5. Limitations, Concerns} \\ \textbf{and Challenges} 
    };
    \node[align=left] at (1, -0.4) { \small{Address some challenges and} \\ \small{measures taken by developers.}};
    \end{scope}

    \begin{scope}[xshift=5cm,yshift = -2.8cm] 
    \draw[fill=gray!10] 
            (-1.2, 1) -- (3, 1) -- (3, -1) -- (-1.2, -1) -- cycle; 
     \node[align=center] at (0.9,0.6){
    \textbf{6. Risks \& Management}
    };
    \node[align=left] at (0.9, -0.3) { \small{Examine S\&P risks explore} \\ \small{some counter measures from} \\ \small{three viewpoints.}};
    \end{scope}

   \begin{scope}[yshift = -2.8cm] 
    \draw[dashed,fill=gray!10] 
            (-1.2, 1) -- (3, 1) -- (3, -1) -- (-1.2, -1) -- cycle; 
    \node[align=center] at (0.9,0.6){
    \textbf{Outro}
    };
    \node[align=left] at (0.9, -0.2) { \small{Carry out a debrief session,}\\ \small{collect additional feedback} \\ \small{and remarks.}};
    \end{scope}

    \draw[-{Stealth[width=3mm,length=4mm]}] 
        (3, 0) -- ++(0.8,0); 

    \draw[-{Stealth[width=3mm,length=4mm]}] 
        (8, 0) -- ++(0.9,0) ;

    \draw[-{Stealth[width=3mm,length=4mm]}] 
        (13, 0) -- ++(0.8,0);
        
    \draw[-{Stealth[width=3mm,length=4mm]}] 
        (16, -1) -- ++(0,-0.8);

    \draw[-{Stealth[width=3mm,length=4mm]}] 
        (13.8, -2.8) -- ++(-0.8,0); 

     \draw[-{Stealth[width=3mm,length=4mm]}] 
        (8.9, -2.8) -- ++(-0.9,0);

     \draw[-{Stealth[width=3mm,length=4mm]}] 
        (3.8, -2.8) -- ++(-0.8,0); 
    
    \end{tikzpicture}
    }
    \caption{Interview Structure.}
    \label{fig_procedure}
\end{figure}

The structure of our semi-structured interviews is outlined below and illustrated in Figure~\ref{fig_procedure}. The interview protocol and other materials are available at our artifact repository. 
The interviews comprised a demographic section and five main sections corresponding to the research questions. Each main section consisted of four to six opening questions, follow-up questions, and, when necessary, additional prompts or explanations. 
\setlength{\abovecaptionskip}{5pt} 

Prior to the interview, all participants were provided with a consent form and a brief project overview.
The project overview outlined the objectives and scope of our study, while the consent form detailed several key considerations and required permissions, such as the consent for audio recording.
Participants were assured of strict data confidentiality, emphasizing that all data used in reporting would be anonymized to prevent the identification of individuals or companies. 
Furthermore, we emphasized that the data would not be shared externally or used in a manner that could be traced back to the participants or their organizations. 
We obtained e-signed consent forms from all participants, confirming they had been granted the necessary permissions for their involvement in the study.
These guidelines were also verbally reiterated before the formal interview commenced.

After addressing any remaining questions and obtaining explicit consent for data handling and recording, we initiated the recording and began the actual interview with the following structure:

\paratitle{1. Demographics.}
The interview begins with collecting background information about the interviewees, such as their development team, positions, work, and AI experience. 

\paratitle{2. Tools and Usage.}
This section will further provide insight into how often AI tools are used in the development process and explore how they are integrated into daily workflow. Specifically, this section will cover AI usage scenarios and their roles in different development stages.

\paratitle{3. Selection and Outcome Evaluation.}
This section examines how interviewees choose and evaluate AI tools, including selection criteria, AI tool comparisons, and how they assess and process AI-generated results, along with their attitudes towards these outcomes.

\paratitle{4. Beneficial and Future Prospects.}
This section examines the interviewees' overall assessment of AI tools and their future expectations. It begins by exploring the differences before and after adopting AI tools, followed by an analysis of their positive impacts. Finally, it looks into their perspectives on the future role of AI tools in their field.

\paratitle{5. Limitations, Challenges, and Concerns.}
This section addresses the limitations, challenges, and concerns interviewees face when using AI tools, including tool limitations, practical bottlenecks, and fears of job displacement by AI. It also explores the measures they take to overcome these challenges and how they adapt their work practices to the changes introduced by AI tools.

\paratitle{6. S\&P Risks and Guidelines.}
This section examines potential security and privacy (S\&P) risks in using AI tools and the guidelines interviewees follow. It covers whether their companies have established guidelines and their attention to risk notifications and disclosures provided by the tools. Lastly, it briefly explores their awareness of relevant international laws and regulations.

Following the structured interview sections, we invited participants to share any additional insights or aspects they felt were overlooked or wished to discuss further. We concluded each interview by expressing gratitude for the participant's valuable time, offering an opportunity for questions and comments, and providing a brief debriefing.

\subsection{Coding and Analysis}
To analyze the interviews, we followed a systematic approach. We digitally recorded and transcribed the audio of each interview using the built-in functions of each meeting platform. 
Each transcript underwent a careful review to correct any transcription errors and ensure accuracy.
For interviews conducted in languages other than English, we utilized DeepL\footnote{\url{https://www.deepl.com/}} for initial translation, followed by a manual review to address potential translation inaccuracies.
To ensure participant confidentiality, we meticulously removed all personally identifiable information from all transcripts.

\noMargine After interviewing 16 participants, we observed that new codes and themes ceased to emerge, with subsequent participants mostly reiterating previously identified themes. Therefore, we decided to stop interviewing after 19 interviews, having reached theoretical saturation. 

Our analytical approach combined inductive and deductive coding strategies, and thematic analysis~\cite{fereday2006demonstrating, swain2018hybrid} as defined by Fereday and Muir-Cochrane. The coding process occurred in two phases, partitioned by Cohen's Kappa inter-rater agreement. 
Initially, when Cohen's Kappa was relatively low and unstable across transcripts, two researchers coded the same transcript independently. 
Once the average Cohen's kappa reached 0.6 (indicating a substantial level of agreement) after the seventh transcript, we transitioned to the secondary phase, where researchers coded different transcripts independently. 
Throughout both phases, we adhered to a consistent coding process: (I) first, we segmented interview transcripts into smaller data units (paragraphs of interviewee's responses); (II) then, two researchers independently coded each transcript using both deductive and inductive approaches. (III) Subsequently, we discussed and refined raw codes (guided by our research questions) and moved them into the main codebook. (IV) We then grouped raw codes into higher-level merged codes (themes), establishing inclusion relationships. (V) Meetings with senior researchers were conducted to deliberate on uncertain codes. 

\subsection{Limitations}
Our study is subject to limitations common to interview-based research, including limited generalizability, potential over- and under-reporting, self-reporting biases, and issues such as recall and social desirability biases~\cite{alshenqeeti2014interviewing, bergen2020everything}. While our sample includes various organizational contexts, from industry-leading companies to startups, it is a convenience sample that may not be fully representative of the broader population of professionals in the software industry. However, this sample still allows us to explore a range of experiences and organizational contexts.


To address the potentially sensitive nature of questions about security and privacy practices, we implemented measures to mitigate social desirability bias. We emphasized to participants that our goal was not to judge their answers but to genuinely understand their approaches and opinions. We also reminded them of their right to skip questions or remove any parts of the interview for any reason. Based on the provided answers, we believe our sample is broad and diverse in geographical representation, professional roles, work experiences, and organizational contexts, ranging from industry-leading companies to startups.

\section{\texorpdfstring{\hyperref[rq:one]{RQ1}}{RQ1}: Adoption and Utility of AI Tools}\label{sec_utility}

In this section, we provide an overview of how AI tools have been adopted and utilized in the software industry. To answer \hyperref[rq:one]{RQ1}, we begin by exploring the diverse types of AI tools currently in use. Following this, we examine the specific scenarios and tasks where software practitioners leverage these AI tools in the software development process.


\subsection{Definition and Selection} 
Although AI tools have become integral to the daily workflows of the software industry~\cite{opara2023chatgpt, greengard2023ai, jungling2019leverage}, our research reveals that practitioners have varying understandings of these concepts. 
These differing perceptions significantly influence how AI tools are utilized in situ. 
We find that participants define and categorize AI tools through multiple dimensions, including specific products, usability features, organizational context, and usage restrictions.

\paratitle{ChatGPT.} ChatGPT and its creator, OpenAI, have emerged as the primary reference points for AI tools in software development. All participants (19) specifically mentioned ChatGPT, often distinguishing between different versions, such as GPT-3.5, GPT-4, and GPT-4o, demonstrating an awareness of the rapid advancements in AI capabilities. As P02 mentioned:

\quoteFrame{``\textit{When we talk about AI tools now, we're usually referring to conversational or large language models like ChatGPT. These conversational tools have been around since last year introduced by OpenAI, and I've been fascinated by them since their release. I've been constantly researching and using them regularly in my work.}''}

\paratitle{GitHub Copilot.} Among AI coding assistant tools, GitHub Copilot stood out as the most widely adopted, with 15 participants mentioning it and 13 actively using it in their daily coding tasks. As P15 noted: ``\textit{I use GitHub Copilot in my daily work for programming. It is a plugin that automatically helps you complete some code. I think it's basically used every day.}'' 

\paratitle{Other AI Tools.} While ChatGPT and GitHub Copilot dominated the discussions, participants mentioned a variety of other AI tools. Gemini and Bing Chat were each used by three participants, while other tools like Claude, Midi Journey, Tabnine, Grok, Dream Maker, ERNIE, Kimi, and Mars Code were mentioned by one or two participants each. This diversity indicates a rich ecosystem of AI tools, each potentially serving specific niches or use cases in software development.


\paratitle{Company Perspectives.} Based on the ownership, participants commonly categorize AI tools into three types: internal AI tools (e.g., Gemini for Google employees), internal non-open-source tools (company-developed AI yet to open to the public), and external AI tools (developed by other companies or individuals outside the company). This categorization reflects the varying degrees of customization and control organizations have over their AI resources, and highlights how companies strategically manage AI deployment to balance innovation, security, and competitive advantage within their development environments.

\paratitle{Usage Restrictions.} While all participants mentioned ChatGPT, only 14 could use it due to various restrictions. Similarly, 15 participants mentioned GitHub Copilot, but only 13 could actively use it. These restrictions could be attributed to factors such as company policies, licensing issues, or regional availability. As mentioned by a participant: ``\textit{The other thing that's different is that you can't use external tools like [external AI tools] inside the company.}'' (P03) Therefore, some participants discussed offline AI tools. P09 explained, ``\textit{I've even used the offline model, which is called Large Language Module studio [...] It is very similar to ChatGPT, if you have some sensitive data you don't want to be seen by others, you can use the offline model.}''

\subsection{Scenarios and Tasks} 

We have explored the factors that affect participants to select and utilize specific AI tools. 
Through interviews, we further discovered that the interviewees employ AI tools across a variety of scenarios and tasks, with their usage patterns largely influenced by their understanding and estimation of AI capabilities for different tasks.

\paratitle{Prescriptive Usages.} When developers have a clear understanding of AI tools' capabilities for specific tasks, they engage in prescriptive usage. This approach is primarily applied to repetitive or routine tasks where AI tools serve as efficient assistants. Coding, a core task for nearly all developers, has become a primary area for AI tool applications. As P04 notes:

\quoteFrame{``\textit{AI tools are integrated into our compiler to help with code completion, error checking, and debugging. It can help me quickly write portions of code, streamlining the development process.}''}

\setlength{\parskip}{1pt} AI tools also significantly contribute to documentation and copywriting tasks. For instance, P03 mentions using AI to expand and polish written content, e.g., ``\textit{When writing documentation, I might draft 30 words and then use AI to expand it to 200 words or to help me polish my writing. As a non-native English speaker, this helps make my writing more authentic and polished.}''

\noMargine Additionally, participants use AI tools such as enhanced search engines for quick information retrieval. P05 explains: ``\textit{With cloud services like AWS offering over 200 different services, it's impossible to remember all the templates and usage of each one. So sometimes, I would use ChatGPT to provide instant answers instead of checking the tech-doc which saves me time.}'' Furthermore, some developers regard AI tools as tutors, turning to them for guidance on unfamiliar or difficult problems.

\paratitle{Continuous Usages.} 
This usage pattern involves ongoing interaction and refinement, particularly for tasks where developers know AI can contribute but may not provide a direct solution directly. 
They use AI tools to obtain real-time information or suggestions, especially for complex or unfamiliar tasks, and then iteratively improve upon the AI's output, which can be regarded as ``in-context Reinforcement Learning from Human Feedback (RLHF)''.
However, as P16 cautions, ``\textit{Directly copy-pasting AI-generated code into your software is dangerous, as you can't be certain of its safety or appropriateness.}'' Consequently, developers typically use AI-generated content as a starting point, engaging in a process of refinement and adaptation to meet their specific needs. This continuous usage reflects developers' understanding of AI tools' strengths and weaknesses, allowing them to leverage AI capabilities effectively while compensating for their limitations through their expertise.

\paratitle{Creative Usages.} When developers are uncertain about AI tools' capabilities in completely solving a task, they engage in creative usage, exploring potential applications and seeking inspiration. 
This approach is particularly evident in ideation and design tasks. As P02 explains:
\quoteFrame{``\textit{Before starting a project, you need to validate your ideas and plans. AI tools like ChatGPT or Kimi can help you explore complementary concepts or check if similar solutions already exist.}''}

\setlength{\parskip}{1pt} In game design, AI tools enhance NPC characterization and prototype design, as P04 describes: ``\textit{For game planning, we use AI to quickly generate creative designs and prototype drawings. It's particularly useful for developing intelligent NPC behaviors, opening up a lot of exploratory possibilities.}''
\find{{\bf \ding{45} Answer to \hyperref[rq:one]{RQ1}:}$\blacktriangleright$ AI tools play multiple important roles throughout the software development process including but not limited to requirements analysis, code generation, and documentation. Most developers use AI as an auxiliary tool. These tools help them efficiently complete repetitive tasks, acting as search engines, tutors, and personal assistants in the development process.}
\section{\texorpdfstring{\hyperref[rq:two]{RQ2}}{RQ2}: Attitudes and Perspectives}\label{sec_attitude}


This section explores software practitioners' views on AI tools in development. To answer \hyperref[rq:two]{RQ2}, we examined AI tools' positive impacts, perceived limitations, and job replacement concerns from participants, providing comprehensive perspectives on AI's role in software development.

\subsection{Positive Impacts} 
The integration of AI tools into software development has yielded significant positive impacts, enhancing various aspects of the development process and revolutionizing how tasks are approached and executed. While the fundamental stages of software development remain consistent, our findings reveal that AI has introduced substantial improvements across the entire development life cycle. These enhancements primarily manifest as increased efficiency and productivity, as highlighted by P06: ``\textit{The main positive impact [of AI tools] is about efficiency, the working efficiency.}'' 

\paratitle{Information Retrieval.} AI has transformed how people access and process information, often surpassing traditional search engines in efficiency, accuracy, and contextual understanding. 
As P08 noted, ``\textit{AI brings you a lot of convenience. Normally, you would search in [search engines] when you find something unknown. Now, you can directly ask AI. The answer AI gives is clearer and more relevant.}'' This enhanced capability extends to knowledge acquisition, where AI tools efficiently filter and summarize vast amounts of online information, providing reliable baselines for learning new technologies. P05 highlighted this advantage: ``\textit{ChatGPT will give you a baseline, [...] you won't be influenced by useless information on the Internet.}'' Moreover, AI demonstrates superior context coherence, maintaining consistency across interactions and effectively handling vague queries. 

\paratitle{Coding.} The coding task has undergone significant transformations with the introduction of AI tools. One of the most notable improvements is the reduction in coding interruptions: ``\textit{Now I can write a dozen lines of code without interruption, whereas before, I might write two lines on average before having to jump out [from development environment] to search engines.}'' (P03) Additionally, these tools have contributed to enhancing code quality and standardization. As P01 explained:

\quoteFrame{``\textit{It allows me to make my code more compliant with software engineering principles, improving usability and readability. With standardized code and refined processes, we ensure better maintainability and adherence to best practices.}''}

\setlength{\parskip}{2pt} In addition, developers can now generate documents more rapidly and express themselves more clearly and professionally. 
As P03 noted, ``\textit{AI can help me generate a large section of the document that I can then modify, rather than starting from scratch.}'' P15 added: ``\textit{When you write a doc [document], you may not know how to express yourself more appropriately, but AI can help you write better.}''
However, certain developers (2) reported that the time saved by using AI for code writing might be negligible in certain scenarios. As P03 mentioned, ``\textit{The time for us to write [code] from scratch and use AI may be the same, but AI makes the work more comfortable by reducing interruptions, leading to a better workflow and improved comfort.}'' 

\paratitle{In-house AI.} Companies are commonly developing tailored in-house AI solutions to facilitate employees on specific tasks. These custom AI tools have shown remarkable potential in enhancing the efficiency of employees' daily tasks. As one interviewee noted: ``\textit{It [the in-house AI] can help us on-call less, reduce the work pressure. When we encounter problems, we can solve them using the AI, and no longer need to contact someone else on-call, waiting for a reply.}'' (P10) 

\paratitle{Quicker Warm-up.} Before AI integration, developers often spent considerable time learning unfamiliar software stacks or content for new projects. As one interviewee described: ``\textit{In the past, when we got a project involving unfamiliar technologies or code structures, we had to learn it from the basics and write demos to test our understanding. This was very time-consuming, sometimes taking a full day just to implement a single function.}'' (P06) However, AI tools have streamlined this process considerably. The same participant noted the shift: ``\textit{With AI, we don't need to fully understand the technologies. You just need to know the required functionality - what input it needs and what output it should produce. You can then instruct the AI to write the code for you. After the AI generates the code, you can quickly test it, and if it works, you're done. This process is much faster.}'' (P06)



\subsection{AI Limitations } 

As AI tools are increasingly integrated into the software development process, it is evident that they also present various challenges and limitations in their usage~\cite{giffari2024analyst,kumar2023current}.
Participants commonly expressed that, when faced with complex problems and abstract concepts, AI often struggles to grasp the context and implicit logical relationships, leading to irrelevant or overly general responses.
These limitations often require users to provide detailed contextual information or manually break down complex tasks into simpler components.
We summarized the limitations from following aspects.

\paratitle{Memorization but not Generalization.} 
The performance of AI tools heavily depends on the pre-trained datasets of their foundational LLMs~\cite{liu2024datasets}, so they exhibit unsatisfying capability on unseen scenarios. 
As P09 pointed out, ``\textit{Because like our latest model, there's some lag in the data it's trained on, and it doesn't train on the latest data.}''
Similarly, P05 said
``\textit{... there will be some problems with the accuracy, because ChatGPT's data doesn't seem to be up to the current day, it seems to be up to last year.}''
Yang et al.~\cite{yang2024unveiling} also conducted research to explore to what extent code models memorize their training data. They found that larger models have more severe memorization issues, especially when generating longer outputs. 
This reliance on memory highlights its lack of generalization ability in practical applications.
When encountering new problems, such as company internal programming languages, existing AI tools often perform poorly. 


\paratitle{Poor Reasoning Capability.} Arkoudas~\cite{arkoudas2023gpt} distinguished reasoning from intelligence, emphasizing that reasoning is more about the process of reaching a conclusion than just the outcome. Some LLMs, however, struggle with effectively applying reasoning skills on relatively complex tasks. This is consistent with our findings, P12 stated that ``\textit{How do you abstract a real problem into an abstract problem, and then do some design in the functional domain based on those abstractions is also important. At this stage, I estimate that AI training will not reach this level in recent years.}'' The first step to developing a new system requires a deep exploration and analysis of real-world problems. This involves not just recognizing surface phenomena, but also delving into the root causes (e.g., locating the root cause of vulnerabilities).


\paratitle{Hallucination.} It is widely acknowledged that hallucination is a major flaw in LLMs. Xu et al.~\cite{xu2024hallucination} has also confirmed that it is impossible to completely eliminate hallucinations in LLMs, as they cannot learn all computable functions.
The dependence on limited training data and a predefined parameter space introduces uncertainty in the output of AI tools. 
When faced with unfamiliar situations or complex reasoning tasks, AI tools may produce false or irrelevant information (e.g., quasi-functional code snippets). 
These problems not only increase the workload for developers but also potentially lead to wasted time, as they need to spend additional time verifying, modifying, or researching information. e.g.,
\quoteFrame{\textit{``Because ChatGPT is not omnipotent, you may ask a very detailed question, [...], then sometimes ChatGPT may not give you a clear answer, you may need to spend a lot of time on GitHub.''} (P05)}

\subsection{Job Replacement Concern} 
The application of AI in software development has sparked significant discussion about its potential impact on job security within the industry~\cite{guardian2023, bessen2018ai, winter2022developers}. Most developers (17) view AI as an assistant to enhance their work rather than a threat, although some participants (4) acknowledged concerns about AI's potential to replace some of their work in the future. Interestingly, five participants shared a philosophical outlook on the potential for job replacement, suggesting developers embrace AI rather than worrying about it: ``\textit{Worrying about AI must have crossed my mind, but it's meaningless. When a new thing comes up, the best approach is to embrace it. [...] It's better to learn to use AI to bring more convenience to our daily lives and make ourselves and our world better.}'' (P14)	

\noMargine While a few participants expressed concerns about potential job displacement, the majority of respondents demonstrated a more optimistic outlook due to AI limitations and new opportunities:
\setlength{\parskip}{0pt} 

\paratitle{AI is Just a Tool.} {A prevalent theme (12) among respondents is the perception of AI as a tool to augment human capabilities instead of a replacement for human developers. AI is seen as enhancing productivity and efficiency in specific tasks: ``\textit{I don't use AI for everything, just for what improves my work efficiency. It's more of a tool, like using a full-featured IDE instead of a basic editor.}'' (P11)}

\paratitle{Complexity of Software Development.} Some participants (7) emphasized that the multifaceted nature of software development poses significant challenges for AI, highlighting two key aspects. First, the current capability of AI tools is limited to some specific tasks in the broader software development process. As P11 noted: ``\textit{There are many aspects of the software development process AI cannot help [...] such as dependency management, security aspects, underlying infrastructure, and CI/CD Pipeline.}'' Second, participants emphasized the necessary understanding of complex business processes across various industries requires human expertise only. 
As P09 elaborated: ``\textit{Many industries, like banking and communications, have unique business processes. Local banks in Europe, for instance, may have different processes than other banks. This aspect needs people to guide the process, which may be inseparable from human factors.}''

\paratitle{New Positions Emergence.} {Some professionals (4) view AI as a catalyst for industry evolution rather than a cause of job losses. They anticipate that while AI may automate certain tasks, it will simultaneously create new roles (e.g., prompt engineer) and opportunities: ``\textit{Even until the day when AI tools can do it [works of software engineer] all, it will definitely create more new jobs. The advent of any tool may eliminate some jobs but also create new opportunities.}'' (P09) }

\paratitle{Reliability and Final Decision-Making.} {Some developers (3) emphasize that AI still falls short in terms of reliability and cannot be trusted for final decision-making. As one participant stated: ``\textit{AI tools are just tools. We don't give them the final say. The final judgment needs to be made by humans.}'' (P12) Another participant added on this point: ``\textit{AI might give good answers to 9.5 out of 10 questions, but the 0.5 questions it gets wrong might be very far off the mark. We definitely need to have some final guarantees, which only humans can provide.}'' (P15)}


\find{{\bf \ding{45} Answer to \hyperref[rq:two]{RQ2}:}$\blacktriangleright$ Practitioners commonly hold a positive attitude towards AI tools. 
AI tools have fundamentally changed the software development workflows, especially the coding tasks, and introduced many positive impacts.
On the other hand, AI tools are also widely criticized due to their poor capability on unseen scenarios, reasoning tasks, and hallucination outputs.
Interestingly, most participants are \textbf{not} worried about their jobs being replaced, as they think AI tools still have significant limitations and new positions will emerge.}
\section{\texorpdfstring{\hyperref[rq:three]{RQ3}}{RQ3}: Security and Privacy Considerations}\label{sec_responsibility}


In this section, we analyze the security and privacy concerns and practitioners' responsibilities related to AI tools in software development. 
To answer \hyperref[rq:three]{RQ3}, we identified potential risks, explored perspectives on risk management responsibilities, and discussed mitigation strategies, providing insights into how the industry stakeholders mitigate and manage these concerns.

\subsection{The Security and Privacy Risks} 

As we delve deeper into the landscape of AI adoption in software development, our analysis shifts from the broader perspectives on AI integration to the specific challenges it presents in terms of security and privacy. In this section, we explored the multifaceted nature of security and privacy risks associated with AI tools in software development from practitioners.

\paratitle{Proprietary Source Code.} Source code is the core digital asset of software service providers.
Software engineers expressed concerns about exposing proprietary source code through the AI tools usage: ``\textit{When integrating AI with our codebase, we can't guarantee that all our code and keys stored in local repositories won't be accessed. If such sensitive information is read by the AI, the potential impact on the company could be substantial.}'' (P07)
P05 also emphasized this risk: ``\textit{[You need] making sure that these things [code] don't upload or leak out, this is a very important constraint of the company.}''
The risk of intellectual property loss extends beyond code to include concepts, algorithms, and innovative approaches. P03 illustrated this concern: ``\textit{If I'm developing a quantitative trading strategy, I'm concerned that my strategy could be read by an AI, become part of its training data, and then someone else could potentially access or replicate it through a prompt.}''

\paratitle{Confidential Data Leakage.} 
Software and datasets are closely interconnected in providing services, but this also raises concerns about sensitive data leakage. The risk extends beyond development-related data to encompass a broader range of confidential information.  P06 explained: ``\textit{We deal with confidential data that can't be made public. If this data becomes public, it could be detrimental to the company and our customers [...] With highly private data, such as government agency information, putting it online could violate the law and lead to serious legal consequences.}''

\paratitle{Potential Copyright Infringement.} As AI systems generate or suggest information (e.g., code) based on their training data, there is a risk of unknowingly incorporating copyrighted or patented materials into one's work. 
This issue becomes even more controversial in creative or design tasks where copyright concerns are more prominent. 
One participant expressed this concern: ``\textit{They [my colleagues] are worried that the model might generate something similar to what it has read before. We might accidentally use something patented by another company, or our own [proprietary code could be exposed].}'' (P11)

\subsection{Perception and Responsibility} 

\noMargine While the above section highlighted concerns around data security, privacy, and intellectual property, participants' responses presented a spectrum of risk awareness and attitudes. Most of the participants (16) acknowledged potential risks but considered them manageable or of limited concern with reasonable countermeasures. A few of the participants (4) expressed a serious consideration of the potential risks when using AI in software development. Interestingly, one participant noted a concerning lack of risk awareness: ``\textit{From last year to this year [2023 to 2024], people actually think very little about the risk of large models. They use it as a toy or tool but never consider that their input information might be leaked.}'' (P13)


The nature of data being processed and the potential consequences of its exposure significantly influence practitioners' risk assessments when using AI tools. For critical business or system data, security concerns often outweigh efficiency gains, as one participant noted: ``\textit{If it involves business operation or system data, it's certainly impossible for us to use external tools to process it. In such cases, safety concerns are bigger than efficiency gains from AI tools.}'' (P12) However, risk tolerance increases for more common or less sensitive tasks. Another developer explained: ``\textit{For groundbreaking work, I'm bound to think about potential risks when using AI tools. But for common tasks like writing a website, it matters less if the AI leaks this to others, because that's what everyone does.}'' (P10) 

The perceived security measures of the AI tool provider influence trust and risk assessment. Developers tend to trust larger, more established AI companies with potentially stronger privacy protection measures, perceiving them as less risky when it comes to data security: ``\textit{... if the app's privacy protection is weak or it belongs to a small company, there may be potential risks of data leakage. When working with AI tools, it's preferable to use apps made by larger companies, as they often provide better security for my privacy.}'' (P08)

The responsibility for managing AI-related risks in software development is perceived to be distributed across various stakeholders, with differing opinions on who should bear this responsibility.

\paratitle{Organizational Responsibility.} Many developers (11) believe risk management should be handled at the company level: ``\textit{Privacy leakage and data security risks in a work context aren't really something that individual Software Development Engineers have to think about. It's an issue the whole company needs to address when formulating policies.}'' (P10), they further elaborated on the psychological impact of such organizational responsibility: ``\textit{When the company takes responsibility for addressing these risks, I don't feel as concerned about their [risks] existence.}'', indicating a potential sense of security or reduced personal concern when risk management is handled at the organizational level.

\paratitle{Specialized Risk Management Team.} Building on the organizational approach, some developers (4) advocate for specialized risk management teams within companies: ``\textit{It's not my personal information that's being compromised, it's at the company level. It's up to the management of the company, or the team that's dedicated to security, to think about how to avoid that at their level.}'' (P11) 

\paratitle{AI Provider Accountability.} Beyond the organizational boundaries, there's a recognition of the role that AI tool providers play in ensuring data security. Some participants (3) highlighted the potential for proactive measures from these providers: ``\textit{When using AI tools, privacy leaks can't be completely avoided. However, AI providers like OpenAI could implement measures to protect user privacy. For instance, they could adjust their model to avoid recording or using data that involves user privacy, while still utilizing non-private data for training future models.}'' (P08)

\phantomsection
\label{re:riskManagement}
\subsection{Risk Management}
We have discussed the potential S\&P risks and responsibilities associated with the use of AI tools.
Furthermore, we observed that both individuals and organizations tend to adopt various measures to mitigate these risks to the greatest extent possible. 
In this section, we will outline the risk protection measures undertaken from three perspectives: individuals, the companies they are affiliated with, and AI tool providers. 

\paratitle{Individuals Perspective.} 
Firstly, developers often implement  \ul{[Information Obfuscation]} processes before transmitting data to AI tools. The most commonly employed methods include data desensitization and anonymization.
\quoteFrame{\textit{ ``Once we're faced with some very private data, like some of the internal data of customs, this kind of private data of government agencies, [...], we will not give to AI under normal circumstances, will only take ABCD, 12345 to replace it.'' (P6)}}

\setlength{\parskip}{2pt} By replacing or removing sensitive data, developers can mitigate the risks of unauthorized access or the leakage of customer information, ensuring that AI only interacts with non-sensitive, carefully filtered data. Additionally, to prevent AI tools from potentially exposing critical business logic or configurations, developers also adopt measures such as removing core logic and essential configuration variables from the code, as supported by P05 \textit{``Try not to copy a lot of code to ChatGPT. You may still have to screen yourself and ask some more basic and detailed questions.''} Before utilizing AI, they ensure that the core logic and crucial configurations are extracted from the code, thereby minimizing potential risks. 

\noMargine Additionally, using  \ul{[Offline AI Tools]} is a practical measurement that is widely adopted by developers, ensuring that sensitive data remains local and is not accessible to external AI providers. 
In the words of P04, \textit{``If the confidentiality is very strong, we will not do it through GPT, because GPT is not our internal product, we will use our own company's product. Then the database will be encrypted and uploaded to our own private library.''}, when dealing with sensitive information, some developers may selectively block AI access to prevent it from interacting with confidential data. According to P09, \textit{``I took some relatively sensitive data to the offline tools when I was doing data analysis before, and it is not allowed to be exposed to the outside. ''}

\paratitle{Company Perspective.} In our interviews, we also found that some developers are not concerned about S\&P risks, as they believe that their companies have already implemented rigorous procedures to ensure the S\&P of using AI tools. 
According to our statistics, only one (1) company has relative  \ul{[Code of Conduct]}, as emphasized by P12 \textit{``It is a code of conduct for all, and it also includes the code of conduct for AI, like the data risk security.}'' These guidelines provide detailed specifications regarding the proper use of AI tools, data security, and the use of communication facilities, ensuring employees adhere to uniform standards when using AI tools. 
For most companies, although formal written guidelines may be lacking, they typically inform employees of relevant precautions through \ul{[Informal Reminders]}, such as verbal reminders, soft rules, or online banners. These reminders often specify which kind of data should not be uploaded to AI tools and which AI tools are permissible for use. As supported by P07: 
\quoteFrame{\textit{``In our development team, it is verbally stipulated, and there is no specific standard for a process. This is not listed, but we have been told verbally, and usually everyone obeys it.''}}

\setlength{\parskip}{2pt} In larger commercial enterprises, there are often dedicated AI development teams or close \ul{[Partnerships with External AI Providers]}. These companies only allow developers to use internal AI tools or those provided by partners in their daily workflow while imposing strict restrictions on external AI tools. The use of external AI tools requires approval, or in some cases, is not permitted. Additionally, companies are extremely rigorous in managing confidential documents and private data. For high-security data, companies implement access and sharing restrictions and establish the \ul{[Security Approval Process]}. Such data is typically only accessible within the company’s intranet or limited to authorized personnel. Companies also restrict personal devices from accessing company resources to ensure that sensitive information is limited to access or leaked without authorization. 

\noMargine Furthermore, many companies emphasize the importance of \ul{[S\&P Training Session]} to increase employees' awareness. For instance, P11 told us that his company provides training courses on how to use Github Copilot to ensure they can follow company regulations in their daily work.
\quoteFrame{\textit{``Since our company working with GitHub Copilot, they have to give the employees an idea of how to use it or what questions people can ask them. That is, to help people solve AI problems, and give people a general idea of how AI could be used.''}}

\vspace{4pt}
\paratitle{AI Tool Providers.}
From the perspective of AI providers, they have implemented a series of measures to alert users to potential risks associated with the use of their AI products and to mitigate these risks. 
One of the most common methods is \ul{[Risk Notification and Disclosure]}. For instance, some companies, like Claude, remind users of potential S\&P risks when they first use the AI tool, ensuring that users remain vigilant throughout their usage. P10 has also observed that \textit{``I'm not actually sure, maybe every Bing chat might have a line underneath it that says something like FYI.}'' (P10) 
AI tools like Bing Chat and ChatGPT explicitly state beneath the text input box that the generated responses are for reference only, to prevent users from relying blindly on the AI's outputs. Similarly, Copilot has previously stated that it does not use user data for training AI models, thereby reducing the risk of improper use of user data, just as P16 stated:
\quoteFrame{\textit{``When Copilot, Microsoft says that it is sensitive. So it will not. So your data stays, and your code stays your code. This is something very important.''}}


\subsection{Awareness of Risk Management and Regulations}

This section explores practitioners' attitudes toward risk management measures and awareness of related regulations of AI tools, which can serve as a guide for improving policy development and compliance within organizations.

\paratitle{Attitude toward Risk Management.}
Participants generally recognize the necessity of risk management at both the company and individual levels, e.g., \textit{``I think it is necessary to affirm the company level, the individual level should also have this awareness, both need to have.}'' (P14) 
They understand the importance of being aware of potential risks and acknowledge effective risk management. However, it is noteworthy that some developers believe that the company's security department should bear more responsibility than individuals. According to P10, ``\textit{The risk of privacy leakage and the risk of data security, when it comes to work, it's not really something that the SDE has to think about, it's really something that the whole company has to think about.''} They argue that the security management department should carefully consider before introducing AI tools. 

\noMargine Developers also have varied opinions on the risk notifications and privacy disclosures of AI tools. The majority of developers (12) do pay less attention to or fail to recognize these risk warnings, often considering them insufficiently prominent. Furthermore, there is significant disagreement among developers regarding the role of these notifications in actual risk mitigation. Some developers believe these notifications still hold some value, fulfilling the ``duty to inform'' and shift responsibility to users in case of issues, and also serve as a reminder and warning to developers. e.g.,
\quoteFrame{\textit{``I think at least they told you in the first time, [...], so you need to be careful when you're using it, and I think it makes sense because when a user, if he gives sensitive information in this situation, I think it's the user's own problem.''} (P06)}

\setlength{\parskip}{2pt} Despite AI tools often claiming that user data will not be used for model training, developers remain doubtful about the credibility of these statements, as users have no way to verify whether their data is indeed not being used. This distrust also stems from the perception that these notifications are more about meeting procedural requirements than serving as effective risk mitigation measures.  

\noMargine Moreover, developers often consider the trade-off between security and efficiency. Just as P12 said \textit{``We internally train a Copilot tool, but it lacks access to a comprehensive external dataset, which limits its effectiveness in certain scenarios. Additionally, the internal model's sample size is smaller compared to open-source alternatives, leading to reduced accuracy and efficiency.}'' While it is necessary to limit the use of AI tools for risk management, such restrictions may impede the productivity and innovation that AI tools can offer.
Specifically, a small number of developers believe that risk notifications, to some extent, hinder AI development. They argue that these notifications may slow down workflows without providing significant security benefits in practical applications.

\paratitle{Consideration of Regulations.}
Privacy and AI-specific regulations play a significant role in the AI-based software development ecosystem~\cite{pan2023toward,pan2024hope,pan2024trap,si2024solution}.
Regarding the understanding of national-level regulations, most developers are familiar with internal company policies, but their comprehension of international regulations is generally limited, with only a few being aware of well-known frameworks such as GDPR~\cite{EU2016} and PDPA~\cite{PDPC}. When adopting AI tools, practitioners rarely consider regulations, and when they do, their focus is primarily on company-level policies. Several factors may contribute to this phenomenon:
1) Some developers consciously avoid regulatory violations by staying within the known boundaries of company policies, leading them to believe no further attention is required. 
2) Many perceive existing regulations as ambiguous, especially regarding AI-specific legal guidance, reducing their inclination to actively consider them. A lack of familiarity with regulations also leads some developers to disregard them altogether.x
3) For developers handling routine tasks without significant privacy risks, regulations may seem unnecessary. 
4) Some believe regulatory compliance should be the responsibility of AI providers or the company's security department, seeing regulations as potential obstacles to efficient development.

\find{{\bf \ding{45} Answer to \hyperref[rq:three]{RQ3}:}$\blacktriangleright$ 
The common S\&P concerns of using AI tools are source code, sensitive data leakage, and potential copyright infringement.
Practitioners hold different opinions about the responsibility toward S\&P risks, ranging from individuals, companies, and tool providers.
Most of the participants consider their companies should take the majority responsibility, and the common risk management measures include [Code of Conduct], [Partnerships with External AI Providers],  [S\&P Training Session], etc.
Moreover, participants commonly have a passive attitude toward risk management and a low awareness of regulations.
}
\section{Lesson Learnt}\label{sec_discuss}
Our study revealed several important insights beyond the primary findings discussed in the results section. These insights provide a broader context for understanding the current state and future trajectory of AI tool adoption in the software development process for different stakeholders.

\subsection{Software Practitioners}
\paratitle{Embrace AI Tools.} Our study reveals that AI tools can significantly enhance productivity and alleviate work pressure in software development. 
Practitioners should embrace these AI tools, viewing them as powerful allies rather than threats to job security. 
Beyond enhancement in daily tasks, such as repetitive tasks, coding, and information retrieval, AI can function as a versatile assistant and tutor, offering intuitive insights across a broad spectrum of topics. It provides developers with a solid foundation for exploration and innovation, acting as a springboard for more advanced problem-solving and creative development.

\paratitle{Master the Art of Prompts.} To fully harness this potential, practitioners must master the art of prompt engineering. This crucial skill involves crafting effective queries and instructions for AI tools, enabling developers to extract more accurate and relevant outputs.
Several specific strategies for prompt engineering are mentioned by participants, such as assigning personas to AI, employing personified methods, continuing iteration, and decomposing complex tasks.

\paratitle{Be Responsible.} However, caution is advised for critical system design or security-sensitive coding tasks. Generally, adhering to company-provided guidelines is advisable, as these typically account for various scenarios and potential risks. In cases where company guidelines are absent, individuals are recommended to be vigilant about the AI tools they employ and the information they input. As discussed in \hyperref[re:riskManagement]{Section 5.4}, common risk management techniques include information obfuscation, following code of conduct, and other precautionary measures.

\subsection{Companies and Organizations}

\paratitle{AI Integration Strategy.} Companies are suggested to proactively integrate AI tools into development environments to maintain competitiveness while implementing robust risk management strategies. We advise involving security departments in the AI adoption process, consulting with AI providers to understand potential vulnerabilities, and collaborating to enhance security measures.
Additionally, the adoption and regulation of AI tools appear to be influenced by organizational factors such as company size and type. 
Larger and established companies, especially in the tech sector, tend to have more structured approaches to AI governance, including learning sessions about security and guidelines for AI use. Such companies are also more likely to invest in developing in-house AI tools. In contrast, startups often allocate fewer resources to regulating AI use, potentially due to resource constraints or a more experimental culture.

\paratitle{Draw the Lines.} Developing a comprehensive code of conduct for AI usage is essential, covering three critical areas. First, acceptable use cases should clearly define when AI tools can and cannot be used. For instance, AI tools might be encouraged for code refactoring or documentation generation but prohibited for critical architectural decisions or handling sensitive data. Second, data handling procedures should establish guidelines on what data can be used as input for AI prompts. This could include classifying data by protection or sensitivity levels. Third, verification processes should be performed to verify AI-generated results before implementation. This could involve automated testing, peer reviews, and establishing criteria for human oversight of AI outputs.

\paratitle{One Size Doesn't Fit.} For companies with specific needs, developing in-house AI solutions can provide tailored assistance while maintaining stricter control over data and processes. In-house AI solutions can take two primary forms: fine-tuned General AI and customized AI tools. Fine-tuned General AI involves customizing existing foundation LLMs with the company's proprietary data and domain-specific knowledge, resulting in an AI assistant that understands the company's unique terminology, processes, and codebase. 
Customized AI tools, on the other hand, are purpose-built AI solutions designed to solve specific tasks within the company's workflow, such as a custom AI code reviewer that understands the company's coding standards and common pitfalls.

\subsection{AI Tool Providers} 

Our findings indicate a strong user preference for secure AI tools. We encourage AI companies to prioritize risk mitigation in their development processes. Besides, looking ahead, AI tool providers could focus on developing solutions that align with the emerging needs and expectations of software practitioners. Based on our findings, we recommend several areas for future development. First, create AI models that integrate knowledge from various disciplines, enabling them to provide more comprehensive assistance in complex software development scenarios. Second, develop seamless integration with popular IDEs, version control systems, and project management tools, ensuring that this integration feels natural and enhances existing workflows rather than disrupting them. Third, respond to the growing demand for on-premises AI solutions that can operate without internet connectivity, addressing data privacy concerns and enabling use in secure environments. Lastly, create specialized AI models for specific industries or programming languages, ensuring a deeper understanding of domain-specific terminologies, best practices, and regulatory requirements.

\subsection{Regulatory Bodies}

\paratitle{Setting Guardrails.} 
Although several pieces of legislation have been introduced regarding AI governance, such as the European AI Act~\cite{madiega2021artificial}, China Basic Security
Requirements for Generative Artificial Intelligence Service~\cite{ChinaBasicRequirementsForGAI}, Singapore Model AI Governance Framework for Generative AI~\cite{SingaporeGAI}.
There is still a clear need for more detailed regulations and implementation guidelines governing AI usage in software development.
We recommend that regulatory bodies develop comprehensive guidelines addressing two critical areas. 
First, guidelines should be established for appropriate AI use cases, particularly concerning customer data and scenarios where AI should not be used to ensure high privacy and security. For instance, regulations might prohibit the use of AI in handling sensitive personal information or in making critical decisions about system architecture without human oversight. 
Second, robust data protection standards should be developed specifically for AI tools in software development, addressing issues such as data anonymization, secure storage of AI training data, and the right to explanation for AI-generated code or decisions.

\paratitle{Stretching with AI.} Regulations should be flexible enough to accommodate rapid technological advancements while providing a robust framework for ethical and secure AI use in software development. 
The controversy surrounding California SB 1047 has sparked widespread discussion among industry professionals, scholars, and researchers~\cite{SB1047}.
To strike the right balance, we suggest implementing a tiered regulatory approach. 
Foundational rules could be established as immutable principles, while more specific guidelines could be regularly reviewed and updated, perhaps on an annual or biennial basis. 
This approach would allow the regulatory framework to evolve in parallel with AI technology. Additionally, we recommend that regulatory bodies invest in developing their own AI expertise, enabling them to make informed decisions and effectively audit AI systems used in software development.

\section{Conclusion}
AI-empowered tools have emerged as a transformative force, fundamentally reshaping the software development industry and promising far-reaching impacts across diverse sectors.
Our study provides the first comprehensive examination of AI tool adoption in software development, offering insights into their utility, developers' attitudes, and associated security and privacy considerations. 
Through 19 in-depth, semi-structured interviews with software practitioners from diverse backgrounds, we explored the behind-the-scenes processes, attitudes, and challenges surrounding AI tool integration in software development workflows.
In conclusion, while AI tools show great promise in revolutionizing software development, their successful integration requires a balanced approach that maximizes their benefits while adequately addressing associated risks and limitations. 
This study contributes to the growing body of knowledge on AI in software engineering and provides a foundation for future research and industry practices in this rapidly evolving field.


\bibliographystyle{ACM-Reference-Format}
\bibliography{11_References}

\end{document}